\documentclass[prd,aps,nofootinbib,floatfix,10 pt]{revtex4}
\usepackage{amssymb}
\usepackage{amsmath,graphicx,color,epsfig}
\usepackage{slashed}
\begin{document}

\title{Parity anomaly in Dirac equation in (1+2) dimensions, quantum electrodynamics and pair production}
\author{Riazuddin}
\email{riazuddin@ncp.edu.pk}
\affiliation{National Centre for Physics, Quaid-i-Azam University Campus, 45320
Islamabad, Pakistan}
\date{\today}
\begin{abstract}

 It is shown that parity operator plays an interesting role in Dirac
equation in (1+2) dimensions and can be used for defining chiral
currents. It is shown that the ``anomalous" current induced by an
external gauge field can be related to the anomalous divergence of
an axial vector current which arises due to quantum radiative
corrections provided by triangular loop Feynman diagrams in analogy
with the corresponding axial anomaly in (1+3) dimensions. It is
shown that the non-conservation of ``chiral charge" due to anomaly
is related with the topological Chern-Simons charge. As an
application pair creation of massless fermions is discussed in
electric field.
\end{abstract}

\maketitle

 The Dirac equation in (1+2) dimensions has some important
features. For one thing it has found application in graphene [1].
Being an equation in odd time-space dimensions, it has some novel
features. For example there exists two inequivalent representations
of Dirac $\gamma$ matrices, without which parity operation and its
conservation would not have been possible. In fact the parity
operator plays an important role and can be used in defining
``chiral" current. Further when an external gauge field is
introduced, it induces an ``anomalous" current which can be related
to the anomalous divergence of an axial current which arises due to
quantum radiative corrections provided by triangular loop diagrams
in analogy with the corresponding axial anomaly in (1+3) dimensions.
This sheds light on the origin of Chern-Simons term. It is shown
that non-conservation of ``chiral charge" due to anomaly is related
to the topological Chern-Simons charge. The use of chiral anomaly in
pair creation of massless fermions in electric field is discussed,
which may be testable.

The Dirac equation for a massless particle in (1+2) dimensions is
[2]
\begin{eqnarray}
i(\gamma ^{\mu }\partial _{\mu })\psi =0
\end{eqnarray}
where
\begin{equation}
\partial _{0}=\frac{1}{c}\frac{\partial }{\partial t}
\end{equation}
[we may replace c by fermi velocity  $v_f$ for a possible
application to graphene] and
\begin{eqnarray}
\slashed\partial = \gamma ^{\mu }\partial _{\mu }=\gamma
^{0}\partial _{0}+\gamma ^{1}\partial _{1}+\gamma ^{2}\partial _{2}
\end{eqnarray}
We will use natural units and put $c = 1$

Now it is known [3] that in 3 space-time dimensions there exists two
inequivalent representations for $\gamma $-matrices (this is true
for any odd number of space-time dimensions), which corresponds to
the choice of sign in $\gamma^0\gamma^1\gamma^2= \pm iI_2$ where
$I_2$ is the two dimensional identity matrix. An explicit matrix
realization of these two representations is given by
\begin{eqnarray}
\gamma ^{0} &=&\sigma ^{3},\text{ }\gamma ^{1}=i\sigma ^{1},\gamma
^{2}=i\sigma ^{2}  \notag \\
\gamma ^{0} &=&\sigma ^{3},\text{ }\gamma ^{1}=i\sigma ^{1},\gamma
^{2}=-i\sigma ^{2}
\end{eqnarray}
Using the first representation Eq. (1) takes the form
\begin{eqnarray}
i\left(\sigma ^3 \partial_0 + i\sigma ^1 \partial_1 + i\sigma ^2
\partial_2 \right)\psi_+ = 0
\end{eqnarray}
One can take the second representation for the Dirac equation at a
space-time point which is obtained by the parity operation:
\begin{eqnarray*}
x \longrightarrow x^p = \left(x^0, x^1, -x^2\right)
\end{eqnarray*}
to obtain
\begin{eqnarray}
i\widetilde{\slashed\partial}\psi_- = 0
\end{eqnarray}
where
\begin{eqnarray}
\widetilde{\slashed\partial } &=&\sigma ^{3}\partial _{0}+i\sigma
^{1}\partial _{1}-i\sigma ^{2}\partial _{2}\notag
\end{eqnarray}
If Eq.(5) is to be invariant under parity, we should be able to
write it as
\begin{equation}
i[\sigma^3 \partial_0^p +i\sigma^1 \partial_1^p + i\sigma^2
\partial_2^p]\psi^p_+(x^p) = 0
\end{equation}
Now since there is no matrix M that simultaneously anti-commute with
$\sigma^2$ and commute with $\sigma^1$ and $\sigma^3$, we can not
find a relation of type
\begin{eqnarray*}
\psi^p_+(x^p) = M\psi_+(x)
\end{eqnarray*}
so as to get Eq. (7) back to Eq. (5). One might therefore argue that
the theory can not be invariant under P. However comparing Eq.(7)
with Eq.(6), namely,
\begin{eqnarray}
i(\sigma^3 \partial_0 + i\sigma^1 \partial_1 - i\sigma^2
\partial_2)\psi_- = 0,
\end{eqnarray}
one sees that a very natural assumption would be that
\begin{eqnarray}
\psi_+^p \left(x^p \right) = -\eta_p \psi_-(x)\notag\\
\psi_-^p \left(x^p \right) = -\eta_p \psi_+(x)
\end{eqnarray}
Thus to preserve parity invariance, we have to take two
representations together; in that case the parity conserving
Lagrangian is
\begin{eqnarray}
\mathcal{L} = \overline \psi_+ (i\partial)\psi_+ + \overline \psi_-
(i\widetilde \partial)\psi_-
\end{eqnarray}
It is convenient to transform to new fields[2]
\begin{eqnarray}
\psi_A &=& \psi_+ \notag \\
\psi_B &=& i\gamma^2 \psi_+
\end{eqnarray}
The Lagrangian (10) can then be written as
\begin{eqnarray}
\mathcal{L}=\overline{\psi }_{A}(i\gamma ^{\mu }\partial _{\mu
})\psi _{A}+ \overline{\psi }_{B}(i\gamma ^{\mu }\partial _{\mu
})\psi _{B}
\end{eqnarray}
It is instructive to put mass term in the Lagrangian (12), which one
can always put equal to zero:
\begin{eqnarray}
\mathcal{L}=\overline{\psi }_{A}(i\gamma ^{\mu }\partial _{\mu
})\psi _{A}+ \overline{\psi }_{B}(i\gamma ^{\mu }\partial _{\mu
})\psi _{B}-m( \overline{\psi }_{A}\psi _{A}-\overline{\psi
}_{B}\psi _{B})
\end{eqnarray}
where under the parity operation
\begin{eqnarray}
\psi _{A,B}^P(x^P)=\eta _P\sigma ^2\psi _{B,A}(x)
\end{eqnarray}
The peculiarity of the parity transformations (14) is that it
changes A states into B states. In fact this doubling of two
components spinors was noted in a pioneering work on gauge theories
in (1+2) dimensions (e.g. QED$_{3}$) [4,5].

 The Hamiltonian density is
\begin{eqnarray}
\mathcal{H}=[\overline{\psi }_{A}(-i\gamma ^{i}\partial _{i})\psi
_{A}+ \overline{\psi }_{B}(-i\gamma ^{i}\partial _{i})\psi
_{B}+m(\overline{ \psi }_{A}\psi _{A}-\overline{\psi }_{B}\psi
_{B})]
\end{eqnarray}
The Hamiltonian density has the so called conjugate symmetry [6],
$\psi_{A}\leftrightarrow \sigma ^{3}\psi _{B},$ in the sense that
$\mathcal{ H\rightarrow -H}$.

It may be noted that the Lagrangian (12) is invariant, even in the
presence of the mass term, under two independent transformations
\begin{eqnarray}
\psi _{A}\rightarrow e^{i\alpha _{A}}\psi _{A}, \text{ }\psi
_{B}\rightarrow e^{i\alpha _{B}}\psi _{B}
\end{eqnarray}
where $\alpha _{A}$ and $\alpha _{B}$ are real, and has thus
$U_{A}(1)\otimes U_{B}(1)$ symmetry. The corresponding conserved
currents are
\begin{eqnarray}
J_{A}^{\mu }=\overline{\psi }_{A}\gamma ^{\mu }\psi _{A},\text{
}J_{B}^{\mu }=\overline{\psi }_{B}\gamma ^{\mu }\psi _{B}
\end{eqnarray}
One can form even (odd) combination corresponding to "vector"
("axial vector") under parity
\begin{eqnarray}
J_{\pm }^{\mu }=1/2[\overline{\psi }_{A}\gamma ^{\mu }\psi _{A}\pm
\overline{ \psi }_{B}\gamma ^{\mu }\psi _{B}]
\end{eqnarray}

An external gauge field $A_{\mu }$ (electromagnetic) can be
introduced by replacing the ordinary derivative by the covariant
derivative ($-e, e> 0$ is the electronic charge, we have put c=1)
\begin{eqnarray*}
\partial _{\mu }\rightarrow D_{\mu }=\partial _{\mu }-ieA_{\mu }
\end{eqnarray*}
So the gauge invariant Lagrangian obtained from Eq. (13) is
\begin{eqnarray}
\mathcal{L}=\overline{\psi }_{A}(i\gamma ^{\mu }D _{\mu })\psi _{A}+
\overline{\psi }_{B}(i\gamma ^{\mu }D _{\mu })\psi
_{B}-m(\overline{\psi}_A\psi_A - \overline{\psi}_B\psi_B)
\end{eqnarray}
This gives the Dirac equation in (1+2) dimensions
\begin{eqnarray}
\lbrack i\gamma ^{\mu }D_{\mu }\mp m]\psi _{A,\text{ }B}=0,
\end{eqnarray}
To get physical insight of Eq. (20) we multiply on left by
$(-i\gamma ^{\nu }D_{\nu }\mp m)$ to put the resulting equation in
the Pauli form
\begin{eqnarray}
\lbrack D^{\mu }D_{\mu }-\frac{e}{2}\sigma ^{\mu \nu }F_{\mu \nu
}+m^{2}]\psi _{A,\text{ }B}=0
\end{eqnarray}
This equation differs from the Klein Gorden equation in the term $e
\sigma ^{\mu \nu }F_{\mu \nu },$ $F_{\mu \nu }=\partial _{\mu
}A_{\upsilon }-\partial _{\nu }A_{\mu }.$ Using
\begin{eqnarray}
\sigma ^{\mu \nu }=\epsilon ^{\lambda \mu \nu }\gamma _{\lambda }
\end{eqnarray}
one can write
\begin{eqnarray}
-\frac{e}{2}\sigma ^{\mu \nu }F_{\mu \nu }=1/2f^{\lambda }\gamma
_{\lambda } \label{20b}
\end{eqnarray}
where $f^{\lambda }$ is current induced by the external gauge field
$A_{\mu }$ [4, 6,7]
\begin{eqnarray}
 f^{\lambda }= -e\epsilon ^{\lambda
\mu \nu }F_{\mu \nu }
\end{eqnarray}
and has abnormal parity. The corresponding induced charge is
\begin{eqnarray}
Q=\int d^{2}x\text{ }f^{0}(x)
\end{eqnarray}
where
\begin{eqnarray}
f^{0}(x)=-e\epsilon ^{ij}F_{ij}=2e(\vec{\nabla}\times
\vec{A})^{3}=2eB
\end{eqnarray}
Here $B$ is the magnetic field perpendicular to the $x-y$ plane.
Thus $Q= 2e\Phi $, where $\Phi $ is the magnetic flux.

The parity conserving Lagrangian which gives Eq. (21) is
\begin{eqnarray}
\mathcal{L} = [\overline{\psi }_{A}(D^{\mu }D_{\mu }+m^{2})\psi
_{A}-\overline{\psi }_{B}(D^{\mu }D_{\mu }+m^{2})\psi
_{B}]+1/2[(\overline{\psi }_{A}\gamma ^{\mu }\psi
_{A}-\overline{\psi }_{B}\gamma ^{\mu }\psi _{B})]f_{\mu }
\end{eqnarray}
so that $f_{\mu }$ is coupled with the current $J_{-}^{\mu }$ given
in Eq. (18) [6]. The Lagrangian is invariant under $U_{A}(1)\otimes
U_{B}(1).$

We now discuss the dynamics responsible for the generation of the
mass introduced above.  We consider a model [8] similar to
Nambu-Jona-Lasino, which has the Lagrangian density invariant under
the chiral transformations (16) and parity:
\begin{eqnarray}
\mathcal{L}=\overline{\psi }_{A}(i\gamma ^{\mu }\partial _{\mu
})\psi _{A}+ \overline{\psi }_{B}(i\gamma ^{\mu }\partial _{\mu
})\psi _{B} + \frac {G}{2}[(\overline \psi_A \psi_A - \overline
\psi_B \psi_B)^2 + (\overline \psi_A \psi_B + \overline \psi_B
\psi_A)^2 + (i(\overline \psi_A \psi_B - \overline \psi_B
\psi_A))^2]
\end{eqnarray}
Note that the 1st term in square bracket is invariant by itself but
the other two terms must come together to preserve the invariance
under the transformations (16) [we now treat $\alpha_A$ and
$\alpha_B$ as infinitesimal]. We now break the chiral symmetry by
giving non-zero expectation values to operators $(\overline \psi_A
\psi_A - \overline \psi_B \psi_B)$ and $(i(\overline \psi_A \psi_B -
\overline \psi_B \psi_A))$ [the parity conservation forbids the
vacuum expectation value of $[(\overline \psi_A \psi_B + \overline
\psi_B \psi_A)$], thereby generating the mass term
\begin{eqnarray}
-m(\overline \psi_A \psi_A - \overline \psi_B \psi_B)
\end{eqnarray}
and the transition mass
\begin{eqnarray}
-m_Ti(\overline \psi_A \psi_B - \overline \psi_B \psi_A)
\end{eqnarray}
Here
\begin{eqnarray}
m &=& -G \langle \overline \psi_A \psi_A - \overline
\psi_B \psi_B \rangle \notag \\
m_T &=& -G \langle i(\overline \psi_A \psi_B - \overline \psi_B \psi_A) \rangle
\end{eqnarray}
It is the term (30) which breaks the chiral symmetry spontaneously.
Then $J_-^\mu$ is no longer conserved and
\begin{eqnarray}
\partial_\mu J_-^\mu = -m_T(\overline \psi_A \psi_B + \overline \psi_B \psi_A)
\end{eqnarray}

Next we discuss whether $f_{\mu }$ can be related to "anomalous"
divergence of some axial current, which arises due to quantum
corrections. Here analogy with the axial vector "anomalous"
divergence in (1+3) dimensions, namely [9]
\begin{eqnarray}
\partial _{\mu }J_{5}^{\mu } &=&-\frac{e^{2}}{32\pi ^{2}}\epsilon ^{\mu \nu
\rho \sigma }F_{\mu \nu }F_{\rho \sigma }  \notag \\
&=&\frac{e^{2}}{4\pi ^{2}}\vec{E}.\vec{B}
\end{eqnarray}
is useful. As is well known this divergence arises from quantum
corrections provided by triangle graph which has two vector vertices
and one axial vector vertex or that provided by its divergence
\begin{eqnarray}
\partial _{\mu }J_{5}^{\mu }=m(\overline{u}i\gamma _{5}u-\overline{d}
i\gamma _{5}d)
\end{eqnarray}
where $u$ and $d$ denote up and down quarks ($m$ is the quark mass),
which provide the internal legs of the triangle. Note that although
quark mass $m$ appears above, but Eq. (33) is independent of quark
mass. Indeed in $(1+2)$ dimensions if we compare Eq. (32) with Eq.
(34) we have corresponding $\psi _{A}$ and $\psi _{B}$ fields, which
appear in the Lagrangian (12) or Hamiltonian (15).

We may remark here that usually an anomaly arises when the quantum
calculation breaks a classical symmetry. In perturbative
calculations this occurs when the regulator breaks the symmetry in
some way. So, for instance, for the triangle anomaly in $(1+3)$
dimesions, the Pauli-Villars regularization breaks chiral invariance
due to mass term for regulator field [reflected also in Eq. (34)].
We may also mention here that if the Lagrangian (19) is replaced by
\begin{eqnarray}
\mathcal{L}=\overline{\psi}(i\gamma^{\mu}D_{\mu})\psi-m\overline{\psi}\psi
\end{eqnarray}
then it has been shown [10] that the mass term and Chern-Simons term
induce each other : $<m\overline{\psi}\psi>=<J^{\mu}>_{m}A_{\mu}$
with
$<J^{\mu}>_{m}=\frac{m}{|m|}\frac{e^{2}}{4\pi}\epsilon^{\mu\nu\rho}F_{\nu\rho}$,
giving rise to parity violation in (1+2) dimensions. However, if one
uses the Lagrangian (19), there is no Chern Simons term since the
sign of fermion mass terms cancel between them any contribution in
this term [11].

 Coming back to our case if $m_T$ were absent there would be no anomaly.
It is the presence of $m_T$ which gives rise to anomaly as we now
show. Noting from Eq. (32) that $\partial _{\mu }J_-^{\mu } $
involves $m_T\overline{\psi}_{A}\psi _{B}$\ and $m_T\overline{\psi }
_{B}\psi _{A},$ the relevant Feynman graphs are shown in Fig 1.

\begin{figure}[th]
\includegraphics[scale=0.6]{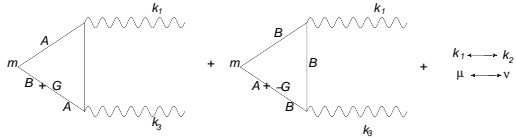}
\caption{Triangle diagrams for "anomalous" current divergence. The
$\times$ denote the vacuum expectation value
$\langle{i(\overline{\psi}_A \psi_B-\overline{\psi}_B
\psi_A)}\rangle$.}
\end{figure}

Note that it is essential to put mass transitions shown (see analogy
with Majorana neutrinos), such mass transitions are provided by
Eq.(30). Noting that A and B propagators involve opposite masses [we
take $m_T = m$], the matrix elements are given by
\begin{eqnarray}
T_{\mu\nu}&=&me^2\int\frac{d^Dl}{(2\pi)^D}\bigg[Tr[\frac{1}{\not{l}+\not{k}_1-m}\gamma_{\mu}\frac{1}{\not{
l}-m}\gamma_{\nu}(\frac{1}{\not{l}-\not{k}_2-m}m\frac{1}{\not{l}-\not{k}_2+m})\notag
\\ &&+\frac{1}{\not {l}+\not{k}_1+m}\gamma_
{\mu}\frac{1}{\not{l}+m}\gamma_{\nu}(\frac{1}{\not{l}-\not{k}_2+m}(-m)\frac{1}{\not{l}-\not{k}_2-m})]\bigg]\notag\\
&=& m^2e^2\int \frac{d^Dl}{(2\pi)^D}\frac{Tr[(\not
{l}-\not{k}_1+m)\gamma_{\mu}(\not{l}+m)\gamma_{\nu}-(\not{l}+\not{k}_1-m)\gamma_{\mu}(\not{l}-m)\gamma_{\nu}]}{[(l+k_1)^2-m^2][l^2-m^2][(l-k_2)^2-m^2]}+\left.
\begin{tabular}{l}
$k_{1}\leftrightarrow k_{2}$ \\
$\mu \leftrightarrow \nu$
\end{tabular} \right.
\end{eqnarray}
The numerator in the integral which contributes is,
\begin{eqnarray}
N^{\mu \nu } &=& 2m\text{ }Tr[\gamma ^{\rho }\gamma ^{\mu }\gamma
^{\nu }(l+k_{1})_{\rho }+\gamma ^{\rho }\gamma ^{\mu }\gamma ^{\nu
}l_{\rho }]
\notag \\
&=&-4m i[\epsilon ^{\rho \mu \nu }(l+k_{1})_{\rho }+\epsilon ^{\mu
\rho
\nu }l_{\rho }]  \notag \\
&=&-4mi\epsilon ^{\mu \nu \rho }k_{1\rho }
\end{eqnarray}

Using the Feynman parametrization, the denominator takes the the
form,
\begin{eqnarray*}
\lbrack (l+k_{1}x-k_{2}y)^{2}-\Delta ]^{3}
\end{eqnarray*}
where
\begin{eqnarray*}
\Delta =m^{2}-xk_{1}^{2}-yk_{2}^{2}+(xk_{1}-yk_{2})^{2}
\end{eqnarray*}
Making the shift $l\rightarrow l-(k_{1}x-k_{2}y)$, the denominator
becomes $ (l^{2}-\Delta )^{3}$ and the dimensional regularization
gives [D=3]
\begin{eqnarray}
T_{\mu \nu } &=&-4m^{3}e^{2}i\epsilon ^{\mu \nu \rho }k_{1\rho
}\int_{0}^{1}dx\int_{0}^{1-x}dy\frac{(-1)^{3}i}{(4\pi
)^{d/2}}\frac{\Gamma
(3-D/2)}{\Gamma (3)}(\frac{1}{\Delta })^{3-D/2}  \notag \\
&&+\left.
\begin{tabular}{l}
$k_{1}\leftrightarrow k_{2}$ \\
$\mu \leftrightarrow \nu$
\end{tabular} \right.
\end{eqnarray}
For photons on the mass shell
\begin{eqnarray*}
k_{1}^{2}=0,\qquad k_{2}^{2}=0
\end{eqnarray*}
and putting $2k_{1}.k_{2}=q^{2},$ $\Delta =m^{2}-q^{2}xy$. Thus we
obtain, neglecting terms of order $q^{2}$,
\begin{eqnarray}
T_{\mu \nu } &=&\frac{e^{2}}{16\pi
}\frac{m^{3}}{(m^{6})^{1/2}}\epsilon ^{\mu \nu \rho }k_{1\rho
}+\left.
\begin{tabular}{l}
$k_{1}\leftrightarrow k_{2}$ \\
$\mu \leftrightarrow \nu $
\end{tabular}
\right.  \notag \\
&=&\frac{e^{2}}{16\pi }(\text{sign }m)\epsilon ^{\mu \nu \rho
}(k_{1}-k_{2})_{\rho }
\end{eqnarray}
Thus finally
\begin{eqnarray}
\partial _{\mu }J_-^{\mu }=\frac{e^{2}}{16\pi }(\text{sign }m)\varepsilon
_{\mu }(k_{1})\varepsilon _{\nu }(k_{2})\epsilon ^{\mu \nu \rho
}(k_{1}-k_{2})_{\rho }
\end{eqnarray}
It may be noted that in (1+3) dimensions $\partial _{\mu }J_5^{\mu
}$ is independent of $m$; in (1+2) dimensions, the corresponding
quantity $\partial _{\mu }J_-^{\mu }$ is also independent of the
magnitude of $m$ but does depend on its sign (which is typical for
odd space-time dimensions). In both cases mass was used as a
regulator. In configuration space Eq. (40) takes the form
\begin{eqnarray}
\partial _{\mu }J_-^{\mu }=\frac{e^{2}}{16\pi }(\text{sign }m)A_{\lambda
}f^{\lambda }=\frac{e^{2}}{16\pi }(\text{sign }m)\epsilon ^{\mu
\nu\lambda} F_{\mu\nu} A_\lambda
\end{eqnarray}
Note that $\epsilon ^{\mu \nu\lambda} F_{\mu\nu} A_\lambda$ is the
Chern-Simons term in (1+2) dimensions. Then for the chiral charge
\begin{eqnarray}
N = N_A - N_B = \int d^2x J_-^0
\end{eqnarray}
we have
\begin{eqnarray}
\Delta N = N(\infty)- N(-\infty) &=& \int_{-\infty}^{\infty} dt \partial_0
\int d^2 x J_-^0(t,\vec x)\notag \\
&=& \int d^3x \partial_\mu J_-^\mu
\end{eqnarray}
where we have used the Gauss's theorem and set the surface integral
equal to zero. Thus
\begin{eqnarray}
\Delta N \neq 0 = \nu,
\end{eqnarray}
where
\begin{eqnarray}
\nu = \frac{e^{2}}{16\pi }(\text{sign }m)\int d^3 x \epsilon ^{\mu
\nu \lambda }F_{\mu \nu }A_{\lambda }
\end{eqnarray}
Note that the integral in Eq. (45) is the Chern-Simons term. Thus
the non-conservation of chiral charge is related with the
topological Chern-Simons charge. In general the topological
connection with Chern-Simons term is discussed in [4,5]

To discuss a possible application of ``anomaly" we write it in terms
of electric and magnetic fields:
\begin{eqnarray}
\epsilon ^{\mu \nu \lambda }F_{\mu \nu }A_{\lambda
}=-2A^{0}B^{3}-2(\vec{A}\times \vec{E})^{3}=-2BA_{0}-B\vec{E}
.\vec{r}
\end{eqnarray}
where B is constant and along z-axis and we have used $\vec A =
\frac{1}{2}\vec B \times \vec r$ so that $\vec B = \vec\nabla \times
\vec A$. One can select a gauge in which $A_0 = 0$ then Eq. (41)
becomes
\begin{eqnarray}
\partial _{\mu }J_{-}^{\mu }=-\frac{e^{2}}{16\pi }(\text{sign }m)B(
\vec{E}.\vec{r})
\end{eqnarray}
[compare it with second line ($\vec{E}.\vec{B}$) of Eq. (33) for
axial anomaly in (1+3)].We now discuss the pair creation of massless
fermions in an axially symmetric electric flux tube $\vec E$ under a
homogeneous magnetic field B perpendicular to the x-y plane in which
the tube lies [12]. Assuming spatial homogeneity of $\vec J_-$ so
that $\vec\nabla . \vec J_- = 0,$ the anomaly  Eq. (46) gives
\begin{eqnarray}
\partial_t(n_B - n_A) = \frac{e^{2}}{16\pi }(\text{sign
}m)\vec{E}(t).\vec{r} B
\end{eqnarray}
where $n_B$ and $n_A$ denote number density of B type and A type
fermions: $\psi^{\dag}_B\psi_B$ and $\psi^{\dag}_A\psi_A$. Further
from $\partial_\mu J_+^\mu = 0$, $\partial_t(n_B + n_A) = 0,$ so
that [sign $m$ is understood]
\begin{eqnarray}
2\partial_t n = \frac{e^{2}}{16\pi }\vec{E}.\vec{r}B
\end{eqnarray}
where we have used $\partial_t n_B = -\partial_t n_A = \partial_t
n$. Now the Fermi momentum $p_F$ is given by
\begin{eqnarray}
p_F &=& \int \frac{d p_F}{dt^\prime}  d{t^\prime} \notag \\
&=& -e\int_{0}^{t} E(t^\prime) dt^\prime
\end{eqnarray}
The energy density of particles is then
\begin{eqnarray}
\epsilon(t) = n(t)p_{F}(t)
\end{eqnarray}
Using the axial symmetry for the cylinderical coordinates $(r, \phi,
z),$ with $\vec B = B\hat z, E = E_r, E_\phi = 0,$ the Maxwell's
equations give
\begin{eqnarray}
\frac{\partial B}{\partial t} = 0, \quad \frac{\partial E}{\partial
t} = -J
\end{eqnarray}
To proceed further, we notice that in relativistic electrodynamics
in (1+2) dimensions, for the action
\begin{eqnarray}
S = \int d^2x dt \mathcal{L}_{em}
\end{eqnarray}
to be dimensionless, $\mathcal{L}_{em}$ has to be [13]
\begin{eqnarray}
\mathcal{L}_{em} = -\frac{1}{2} ({\vec B}^2 -{\vec E}^2)^\frac{3}{4}
\end{eqnarray}
the dimensions (in natural units) of $A^\mu, \vec B$ and $\vec E$
being respectively $L^{-1}, L^{-2}, L^{-2}.$ The electromagnetic
energy density is then given by
\begin{eqnarray}
H_{em} = \frac{1}{2} (B^2+\frac{1}{2} E^2)(B^2-E^2)^{-\frac{1}{4}}
\end{eqnarray}
which has the right dimensions. The energy conservation then gives
[using Eqs. (52)]
\begin{eqnarray*}
\frac {\partial}{\partial t} \int d^2x [\frac{1}{2} (B^2+\frac{1}{2}
E^2)(B^2-E^2)^{-\frac{1}{4}} + \epsilon]~~~~~~~~~~ \notag \\ = \int d^2x
[-\frac{3}{4} JE(B^2-\frac{1}{2} E^2)(B^2-E^2)^{-\frac{5}{4}} +
\frac {\partial\epsilon}{\partial t}] = 0
\end{eqnarray*}
We will now assume that $B > > E$, then
\begin{eqnarray}
\int d^2x[-\frac{3}{4} JE B^{-\frac{1}{2}} + \frac {\partial
\epsilon}{\partial t}] = 0
\end{eqnarray}
But from Eqs. (49), (50) and (51)
\begin{eqnarray}
\frac{\partial\epsilon}{\partial t} &=& \partial_t n(t) p_F(t) +
n(t)\partial_t p_F(t)\notag \\
&=& \frac{e^2}{32\pi} rBp_F - n(t)eE(t)
\end{eqnarray}
so that from Eq. (56)
\begin{eqnarray*}
\int d^2x E[-\frac{3}{4} J B^{-\frac{1}{2}} + \frac{e^2}{32\pi}
rBp_F - en(t)] = 0
\end{eqnarray*}
implying
\begin{eqnarray}
J(r, t) = \frac{4}{3} B^{\frac{1}{2}} [-e n(r, t) +
\frac{e^2}{32\pi} rBp_F]
\end{eqnarray}
The use of Eqs. (49) and (50) allows us to write
\begin{eqnarray}
\frac{e^2}{32\pi} rBp_F &=& \frac{e^2}{32\pi} \int_{0}^{t}
dt^\prime(-e) E(r, t^\prime)r B\notag \\ &=& -e\int_{0}^{t}
dt^{\prime}\partial_{t^\prime} n(r, t^\prime)\notag \\ &=& -e n(r,
t)
\end{eqnarray}
with $n(r, 0) = 0$. Thus finally Eq. (58) gives
\begin{eqnarray}
J(r, t) = -\frac{8}{3} eB^{\frac{1}{2}}n(r, t),
\end{eqnarray}
so that
\begin{eqnarray}
\frac{\partial E}{\partial t} = \frac{8}{3} eB^{\frac{1}{2}} n(r, t)
\end{eqnarray}
We have to solve the Eqs. (49) and (61), which give
\begin{eqnarray*}
\frac{\partial^2 E}{\partial^2 t} = \frac{e\alpha}{3}
B^{\frac{3}{2}}Er
\end{eqnarray*}
where $\alpha = \frac{e^2}{4\pi}$, the fine structure constant. This
has the solution
\begin{eqnarray}
E(r, t) = E_0(r)\cos\left(\sqrt{\frac{e\alpha
B^{\frac{3}{2}}r}{3}}t\right)
\end{eqnarray}
and then from Eq. (61),
\begin{eqnarray}
en(r, t)= \frac{3}{8}E_0(r) \frac{1}{B^{\frac{1}{2}}}(\frac{e\alpha
B^{\frac{3}{2}}r}{3})^{\frac{1}{2}}
\left|\sin\left(\sqrt{\frac{e\alpha
B^{\frac{3}{2}}r}{3}}t\right)\right|
\end{eqnarray}
We note that even we if start with spatially independent electric
field $E_0$ at $t=0$, it develops $r$-dependence subsequently due to
the anomaly equation (49). This is the main difference from (1+3)
dimensions [12]. The natural length in the system is the magnetic
length $l_B = (\frac{1}{eB})^{\frac{1}{2}}$, which also appears in
the Landau levels. Then [we take $E_0$ independent of $r$]

\begin{figure*}[ht]
\begin{tabular}{cc}
\hspace{0.6cm}($\mathbf{a}$)&\hspace{1.2cm}($\mathbf{b}$)\\
\includegraphics[scale=0.5]{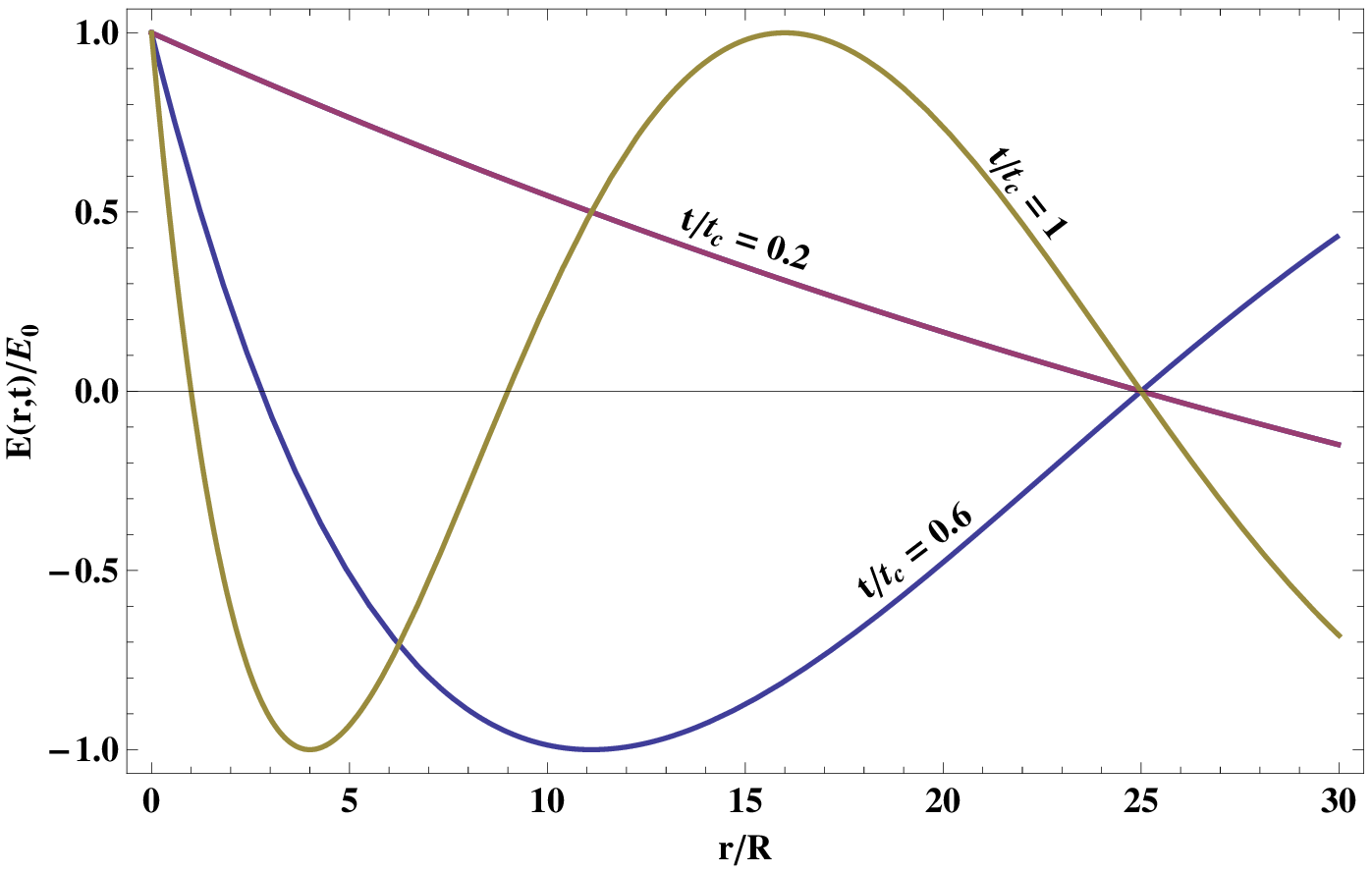} \ \ \
& \ \ \ \includegraphics[scale=0.5]{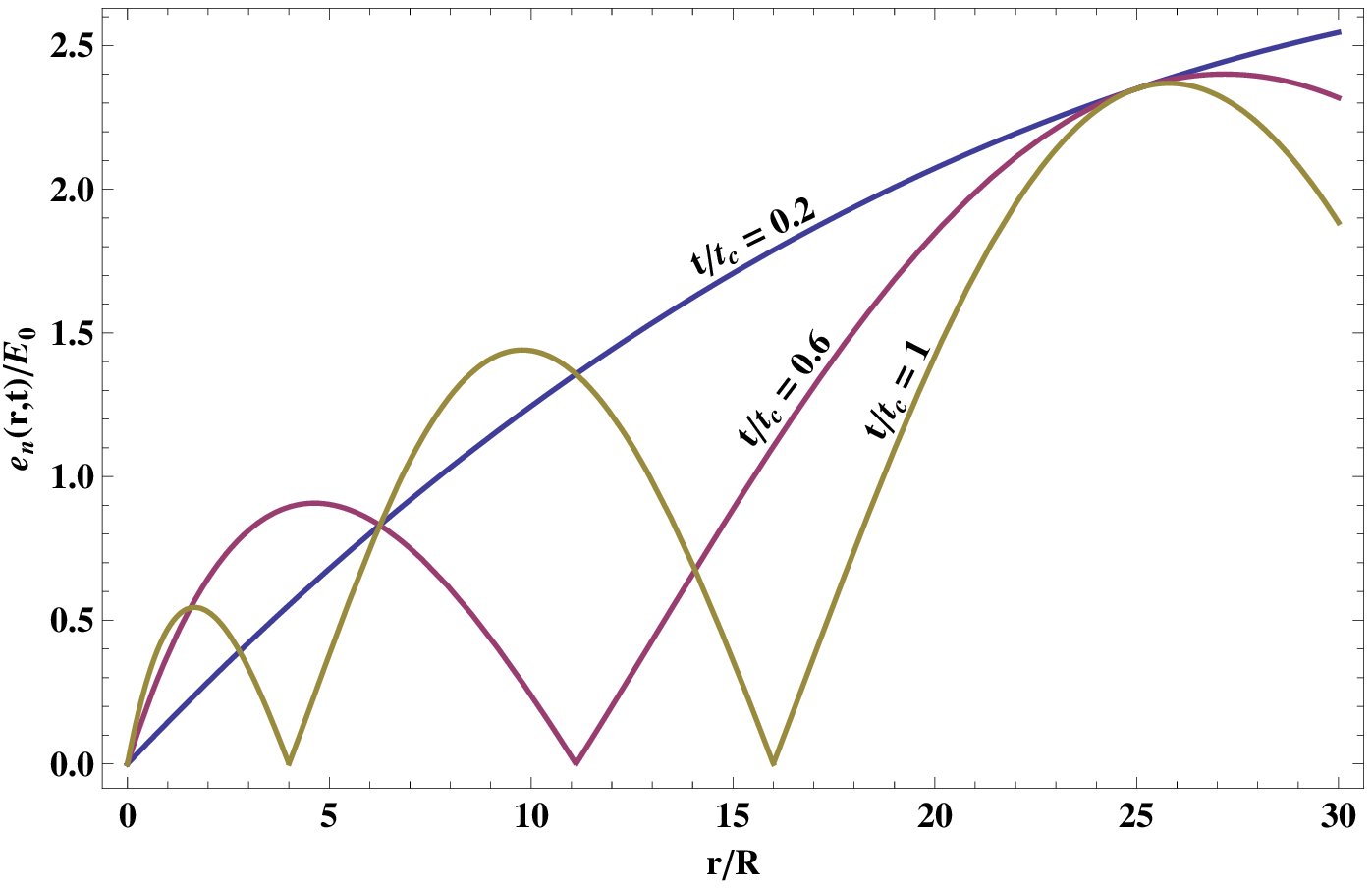}
\end{tabular}
\caption{(a) Electric field $E(r)/E_0$ at $t/t_c=$0.2, 0.6 and 1. (b) $e_n(r)/E_0$ at $t/t_c=$0.2, 0.6 and 1. } \label{lpm}
\end{figure*}

\begin{eqnarray}
\frac{E(r, t)}{E_0} =
\cos\left(\sqrt{\frac{\alpha}{3e^{\frac{1}{2}}} \frac{r}{l_B}}
\frac{t}{t_B}\right)
\end{eqnarray}
and
\begin{eqnarray}
\frac{en(r, t)}{E_0} = \frac{3}{8}(\frac{\alpha}{3e^{\frac{1}{2}}}
\frac{r}{l_B})^{\frac{1}{2}}
\left|\sin(\sqrt{({\frac{\alpha}{3e^{\frac{1}{2}}}\frac{r}{l_B}})}\frac{t}{t_B}\right|
\end{eqnarray}
where $t_B = (\frac{1}{eB})^{\frac{1}{2}}=l_B$ in our units.

We can see from Eqs. (64) and (65) that $E(r, t)$ and $n(r, t)$
oscillate with time for given $r$. If the particles are confined
within the radius $R (R>>l_B)$ such that $E(R, t_C) = 0$, where
$t_C$ may be roughly defined for the life time of the electric field
[12], so that $t_C = \frac{\pi}{2}t_B(\frac{\alpha
R}{{3e^{\frac{1}{2}}}l_B})$. Then we can write Eqs. (64)and (65) as
follows
\begin{eqnarray}
\frac{E(r, t)}{E_0} &=& \cos(\frac{\pi}{2}(\frac{r}{R})^{\frac{1}{2}}\frac{t}{t_C})\notag \\
\frac{en(r, t)}{E_0} &=&
\frac{3}{8}(\frac{\pi}{2}\frac{r}{R})^{\frac{1}{2}}\left|\sin(\frac{\pi}{2}(\frac{r}{R})^{\frac{1}{2}}\frac{t}{t_C})\right|
\end{eqnarray}
These are plotted in Figs. 2a and 2b for $\frac {t}{t_C} = 0.2, 0.6$
and 1. The behavior shown in figs (2a) and (2b) may be testable and
might find applications in quark gluon plasma when E and B are
regarded color electric and magnetic fields in going from QED to QCD
[12]. The above considerations might have applications in graphene
[13] [where c is to be replaced by $v_f$].

The author would like to thank Ansar Fayyazuddin for some useful
comments. He would also like to thank S.Deser for pointing out
ref.[4] and R.Jakiw for pointing out the refs.[4,5,10].


\begin{thebibliography}{9}

\bibitem{1} A.H.Castro Neto, G.Guinea, N.M.R. Peres, K.S.Novoselov, and A.K.
Geim, Rev. Mod. Phys. 81, 109 (2007).

\bibitem{2} For a review,see Riazuddin, IJMPB,26, 1242005 (2012),
arXiV:0807.4804.

\bibitem{3} See for a review, lecture notes of a short course given on
\textquotedblleft Origin of Mass\textquotedblright\ by Adnan Bashir
at National Centre for Physics, Quaid-I-Azam University, Islamabad,
in Dec. 2005; see also A. Raya and E. D. Reyes, arxiv: 0807.0829.

\bibitem{Deser} S.Deser, R. Jackiw and S.Templeton, Ann.Phys.(NY)
140,372 (1982).

\bibitem{GD} For a review see G.V.Dunne, arXiv: hep-th/9902115.

\bibitem{4} G. W. Semenoff, Phys. Rev. Lett. 53, 2449 (1984).

\bibitem{6} R. Jackiw, Phys. Rev. D 29, 2375 (1984).

\bibitem{5} I. A. Shovkovy, arxiv: 1108.4656.

\bibitem{7} S. Alder, Phys. Rev. 117, 2426 (1969); J. S. Bell and R.
Jackiw, Nuovo Cimento, 60A, 47 (1969).

\bibitem{ANR} A.N.Redlich, Phys.Rev.Lett.52,18(1984)  and
Phys.Rev.D 29, 2366(1984); For a review see ref. \cite{GD}.

\bibitem{AR} A.Raya and E.D.Reyes ref \cite{3}.

\bibitem{8} A. Iwazaki, arXiv: 0905.2003.

\bibitem{9} M. I. Katsnelson and G.E. Volovik, Pis'ma ZhETF 95, 457-461 (2012), arxiv:
1203.1578.
\end{thebibliography}
\end{document}